\documentclass[submission,copyright,creativecommons]{eptcs}
\usepackage{breakurl}             
\usepackage{threeparttable}
\usepackage{amssymb}
\usepackage{graphicx}
\usepackage{color}
\usepackage{comment}
\usepackage{url} 
\usepackage{multirow}
\usepackage{amsmath}
\usepackage{xspace}
\usepackage{subfigure}

\newcommand{\G}{\textbf{G}\xspace}
\newcommand{\F}{\textbf{F}\xspace}
\newcommand{\X}{\textbf{X}\xspace}

\newcommand{\sig}[1]{\textsf{#1}\xspace}

\title{Numerical LTL Synthesis for Cyber-Physical Systems \\\begin{large}(Work-in-Progress Report)\end{large}}
\author{Chih-Hong Cheng
\institute{fortiss GmbH}
\email{cheng@fortiss.org}
\and
Edward A. Lee
\institute{EECS, UC Berkeley}
\email{eal@eecs.berkeley.edu}
}
\def\titlerunning{Numerical LTL Synthesis for Cyber-Physical Systems}
\def\authorrunning{Chih-Hong Cheng, Edward A. Lee} 
\begin{document}
\maketitle

\begin{abstract}
\end{abstract}

\section{Overview}

Cyber-physical systems (CPS) are integrations of computation with physical processes, where computing machineries monitor and interact with the physical world via sensors and actuators. In such a system, the reading of a sensor represents measurements of a physical quantity, and the values are often reals ranged over bounded intervals. The implementation of control laws is based on nonlinear numerical
computations over the received sensor values. Synthesizing controllers fulfilling features within CPS brings
a huge challenge to the research community in formal methods, as most of the work in automatic controller
synthesis (LTL synthesis) is designed to handle specifications having inputs within the Boolean domain~\cite{Jobstm06c,ScheweF07a,Ehlers11,acacia12}.

In this report, we present a novel approach that addresses the above challenge to synthesize controllers for CPS. Our core methodology, called \emph{numerical LTL synthesis}, extends LTL synthesis~\cite{pnueli1989synthesis} by using inputs or outputs in real numbers and by allowing predicates of polynomial constraints to be defined within an LTL formula as a specification. The synthesis algorithm is based on an interplay between an LTL synthesis engine that handles the pseudo-Boolean structure, together with a nonlinear constraint validity checker that tests the (in)feasibility of a (counter-)strategy. The methodology is integrated within the CPS research framework Ptolemy~II~\cite{eker2003taming} via the development of an LTL synthesis module \textsf{G4LTL}~\cite{g4ltl} and a validity checker \textsf{JBernstein}~\cite{cheng:JBernstein}. Although we only target the theory of nonlinear real arithmetic, the use of pseudo-Boolean synthesis framework also allows an easy extension to embed a richer set of theories, making the technique applicable to a much broader audience.

\section{Numerical LTL Synthesis}

\subsection{Understanding Numerical LTL Synthesis}
We use an example to illustrate key ingredients of numerical LTL synthesis, including the specification methodology and the algorithmic interplay between the pseudo-Boolean LTL synthesizer and the theory solver.

\paragraph{Specification.} Consider a reactive system where in each reactive cycle, sensor readings are updated. In this system, two variables $\sig{x}, \sig{y}\in [0, 4] \cap \mathbb{R}$ are used to store the sensor values in each cycle. The system has two clients \textsf{Client1} and \textsf{Client2}. \textsf{Client1} issues a request when $\sig{x}+\sig{y}>3$ holds, while \textsf{Client2} issues a request when $\sig{x}^2+\sig{y}^2<\frac{7}{2}$ holds. Our goal is to design a controller that grants a client in the next round whenever he requests. However, as the resource is unique, it is disallowed to grant the resource simultaneously to two clients in each cycle. When we use two Boolean variables \sig{grant1}, \sig{grant2} to represent the issuing of the resource in each round, we can describe the specification of a correct controller formally using the following ``extended" LTL formula, where \G, \F and \X are temporal operators~\cite{pnueli1977temporal} representing ``always," ``eventually" and ``next".

\begin{equation} \label{eq.original.spec}
 \G (\sig{x}+\sig{y}>3 \rightarrow \X\, \sig{grant1})
 \wedge \G (\sig{x}^2+\sig{y}^2<\frac{7}{2} \rightarrow \X\, \sig{grant2}) \wedge
 \G (\neg(\sig{grant1} \wedge \sig{grant2}))
\end{equation}

\paragraph{Pseudo-Boolean specification synthesis.} The implemented numerical LTL synthesis algorithm interleaves between the pseudo-Boolean level and the theory level. First, rewrite $\sig{x}+\sig{y}>3$ and $\sig{x}^2+\sig{y}^2<\frac{7}{2}$ as predicates $\sig{req1}$ and $\sig{req2}$. View $\sig{req1}$ and $\sig{req2}$ as independent input variables, then we have created the following pseudo-Boolean specification.

\begin{equation}\label{eq.abstract.spec}
 \G (\sig{req1} \rightarrow \X\, \sig{grant1})
 \wedge \G (\sig{req2} \rightarrow \X\, \sig{grant2}) \wedge
 \G (\neg(\sig{grant1} \wedge \sig{grant2}))
\end{equation}

In the original specification, valuations of $\sig{x}+\sig{y}>3$ and $\sig{x}^2+\sig{y}^2<\frac{7}{2}$ are inter-related, but \sig{req1} and \sig{req2} in the new specification are independent input variables.
Therefore, the pseudo-Boolean specification \emph{overapproximates} the capability of the environment.

When an LTL synthesis engine analyzes Eq.~\ref{eq.abstract.spec}, it reports the unrealizability of the specification. The counter-strategy (i.e., reason of unrealizability) can be understood by a simple input pattern $(\sig{true}, \sig{true})$, as when both $\sig{req1}$ and $\sig{req2}$ are \sig{true}, a controller needs to assign \sig{grant1} and \sig{grant2} to be \sig{true} in the next round. This violates the third conjuncted invariance condition $\G (\neg(\sig{grant1} \wedge \sig{grant2}))$.

\paragraph{Counter-strategy validation (theory level).} Given a counter-strategy on the pseudo-Boolean level, one needs to check whether such a strategy is indeed possible. In this example, it is equivalent to check whether it is possible to have $\sig{x}+\sig{y}>3$ and $\sig{x}^2+\sig{y}^2<\frac{7}{2}$ true simultaneously, given $\sig{x}, \sig{y} \in [0, 4]$. We use a validity checker to examine the negated constraint $\forall \sig{x}, \sig{y}\in [0, 4]: \neg(\sig{x}+\sig{y}>3 \wedge \sig{x}^2+\sig{y}^2<\frac{7}{2})$, or equivalently:

\begin{equation}\label{eq.validity}
\forall \sig{x}, \sig{y}\in [0, 4]: \sig{x}+\sig{y}>3 \rightarrow \sig{x}^2+\sig{y}^2\geq\frac{7}{2}
\end{equation}

Such an assume-guarantee style constraint can be checked automatically by the solver \textsf{JBernstein}. In this example, \textsf{JBernstein} returns \textsf{true}, implying that this counter-strategy is spurious. If all pseudo-Boolean input patterns in the counter-strategy are realizable on the theory level, then the counter-strategy is genuine.


\paragraph{Specification refinement and resynthesis (pseudo-Boolean level).} As the input pattern (\textsf{true}, \textsf{true}) can never appear, the subsequent step on the pseudo-Boolean level is to perform specification refinement. To rule out (\textsf{true}, \textsf{true}), one adds an \emph{assumption} $\G (\neg(\sig{req1} \wedge \sig{req2}))$ to the original specification.

\vspace{-2mm}
\begin{small}
\begin{multline}\label{eq.abstract.spec.refined}
 \G (\neg(\sig{req1} \wedge \sig{req2})) \rightarrow ( \G (\sig{req1} \rightarrow \X\, \sig{grant1})
 \wedge \G (\sig{req2} \rightarrow \X\, \sig{grant2}) \wedge \G (\neg(\sig{grant1} \wedge \sig{grant2})))
\end{multline}
\end{small}
\vspace{-2mm}

By posing this assumption and re-running the LTL synthesis engine, the engine reports the existence of a solution: if any one of \sig{req1} ($\sig{x}+\sig{y}>3$) and \sig{req2} ($\sig{x}^2+\sig{y}^2<\frac{7}{2}$) is \textsf{true}, grant the corresponding client in the next round.


\begin{figure}[t]
	\centering
	\includegraphics[width=0.55\columnwidth]{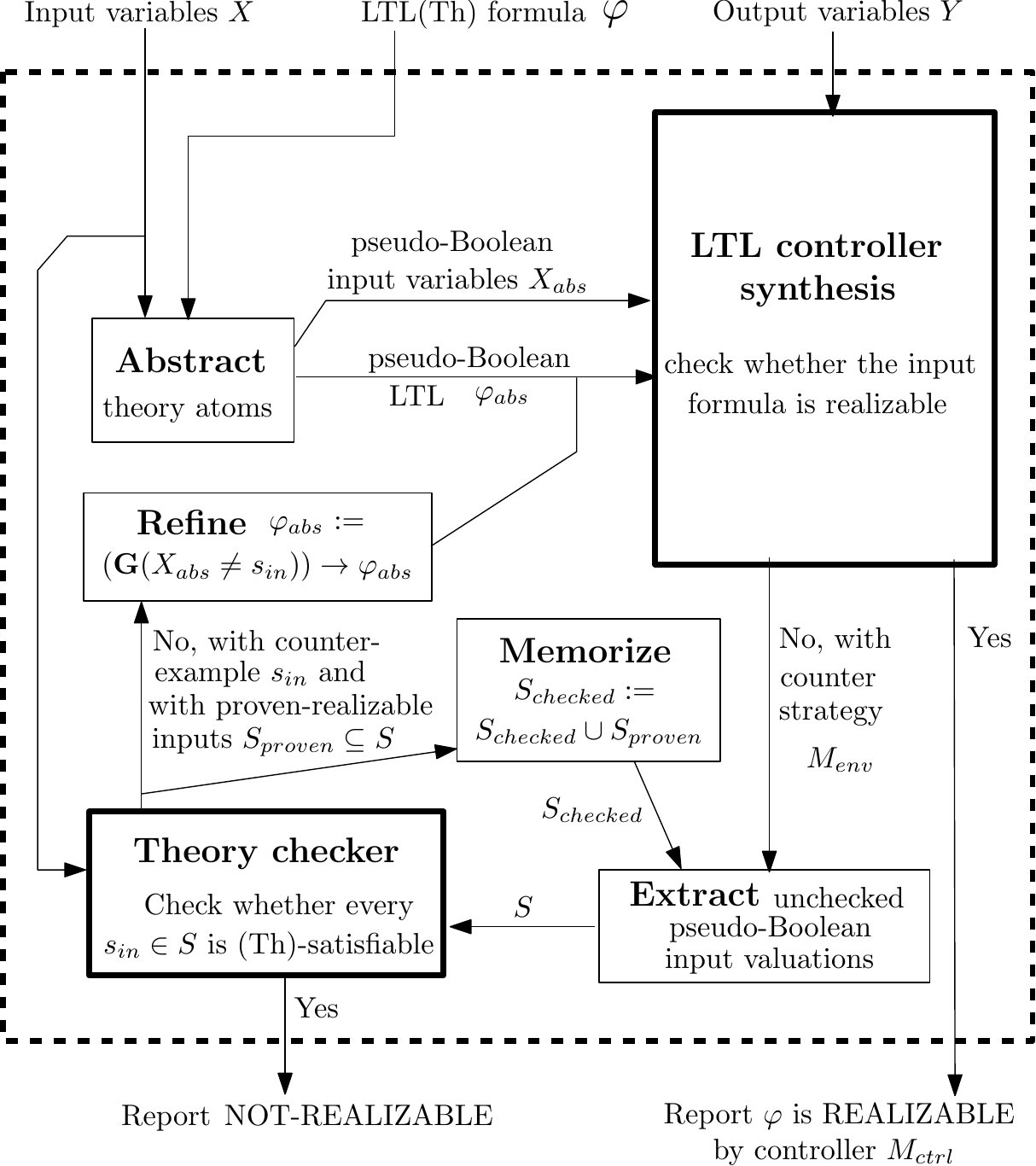}
	\caption{The workflow for CEGAR-based numerical LTL synthesis.}
	\label{fig:CEGAR}
\end{figure}

\subsection{Workflow}

The whole workflow for numerical LTL synthesis is illustrated in Fig.~\ref{fig:CEGAR}. The input of the workflow contains \textsf{LTL(Th)}, the extended LTL formula where \textsf{Th} is the underlying theory. For the presented example, \textsf{Th} is the theory of nonlinear real arithmetic.  The workflow first creates the pseudo-Boolean specification via the \textsf{Abstract} module, then the pseudo-Boolean specification is fed into the LTL controller synthesis engine. If the engine finds a controller $M_{ctrl}$, then the workflow stops and reports realizability. Otherwise, the workflow generates  $M_{env}$, the counter-strategy for the environment to act as a spoiler\footnote{LTL synthesis is known to be \emph{determined}: If one can not find a controller $M_{ctrl}$ realizing the LTL specification, there exists a counter-strategy $M_{env}$ for the environment.}.

Given $M_{env}$, the \textsf{Extract} module extracts a compact strategy that uses fewer pseudo-Boolean input valuations whose validity are not proven on the theory level. It then passes these unproven input valuations $S$ to the theory checker. In the previous example, the \textsf{Extract} module generates a counter-strategy that only uses $(\sig{req1}, \sig{req2})=(\sig{true}, \sig{true})$, and $S=\{(\sig{true}, \sig{true})\}$. Given $S$, the theory checker checks whether any element in $S$ is not realizable. If all of them are realizable, then the counter-strategy is genuine and the workflow stops by reporting the unrealizability of the specification. Otherwise, whenever the theory checker detects one input combination $s_{in}$ that is not realizable, $s_{in}$ is used to refine the pseudo-Boolean specification. The refinement is demonstrated in the \textsf{Refine} module, and the refined specification is fed again to the LTL synthesis engine. For the set $S_{proven}$ of pseudo-Boolean input combinations that are proven to be realizable by the theory checker, it is stored within the \textsf{Memorize} module as elements in $S_{checked}$ (initially, $S_{checked} = \emptyset$), so that later in the input extraction process, module \textsf{Extract} does not place any element inside  $S_{checked}$ and avoids duplicate checking on the theory level.

Our proposed algorithm can be viewed as analogous to the counter-example guided abstraction refinement (CEGAR)
techniques~\cite{clarke2000counterexample} applied in program verification. An intuitive method to
replace the above procedure is to check whether all possible input combinations are possible within the
theory level. However, this can be overly expensive, as given $n$ pseudo-Boolean input variables,
one needs to perform validity checking of nonlinear constraints for $2^n$ times. Our counter-example
guided technique is superior, as demonstrated in the previous example, it is sufficient to trigger the validity
checking once rather than four times. Also, given the infeasibility of a certain input vector on the
pseudo-Boolean level, it is impossible to have a counter-strategy that reuses this input vector, as
this makes the specification trivially true. Together with the use of the \textsf{Memorize} module, the workflow guarantees
that every input vector on the pseudo-Boolean level is checked for its validity at most once.

\vspace{-2mm}
\subsection{Duality between Inputs and Outputs}

Clearly, the above problem formulation and workflow are simplified, as they only consider
output variables within the Boolean domain. The methodology can be naturally extended so that
the every output variable is also replaced by a polynomial constraint\footnote{Here we pose a restriction
on each polynomial constraint such that it can not be mixed with both input and output variables. Without the restriction,
in the implementation one needs an analysis tool for exists-forall formulas.}. In this way, even if a controller $M_{ctrl}$ is
successfully synthesized on the pseudo-Boolean level, one still needs to check whether every output of $M_{ctrl}$ is
realizable, similar to the above input validation process for counter-strategies. For pseudo-Boolean outputs,
the refinement of un-realizability is to add new constraints as \emph{guarantees} in the pseudo-Boolean specification.
E.g., let \sig{grant1} and \sig{grant2} be output predicates on the pseudo-Boolean level. If we
derive that \sig{grant1} and \sig{grant2} can not appear true simultaneously as a fact from the theory
level, add $\G (\neg(\sig{grant1} \wedge \sig{grant2}))$ to refine the pseudo-Boolean specification.



\section{Towards an Optimized Implementation in Ptolemy II}

\begin{figure}[t]
\begin{scriptsize}
\begin{subfigure}

        \begin{verbatim}
## Assume that the operator always comes once for a while
ASSUME ALWAYS (EVENTUALLY (operator))

## Whenever the error sign is raised,
## stop the process until the operator comes
ALWAYS (error -> (stop UNTIL operator))

## Rewrite <->, as it is not supported by LTL2Buchi
ALWAYS (stop -> (!grant1 && !grant2 && !grant3) )
ALWAYS ((!grant1 && !grant2 && !grant3) -> stop)

ALWAYS (req1 -> EVENTUALLY grant1)
ALWAYS (req2 -> EVENTUALLY grant2)
ALWAYS (req3 -> EVENTUALLY grant3)
ALWAYS (!grant1 || !grant2)
ALWAYS (!grant2 || !grant3)
ALWAYS (!grant1 || !grant3)

INPUT error, operator, req1, req2, req3
OUTPUT stop, grant1, grant2, grant3
        \end{verbatim}
\end{subfigure}
\vspace{-3mm}
\begin{subfigure}

\begin{verbatim}
## Newly translated specification
ASSUME ALWAYS (EVENTUALLY (operator))

ALWAYS (error -> ((!sig1 && !sig2) UNTIL operator))
ALWAYS (req1 -> EVENTUALLY (!sig1 && sig2))
ALWAYS (req2 -> EVENTUALLY (sig1 && !sig2))
ALWAYS (req3 -> EVENTUALLY (sig1 && sig2))

INPUT error, operator, req1, req2, req3
OUTPUT sig1, sig2
\end{verbatim}
\end{subfigure}
         \end{scriptsize}
	\caption{An example for specifying a controller with error handling capabilities (above),
            and the translated specification by \textsf{G4LTL} (below).}
	\label{fig:Error.Handling}
\end{figure}

In this section, we outline how we integrate the presented numerical LTL synthesis technique into Ptolemy~II. Ptolemy~II is an open source CPS research framework which is implemented in Java and supports easy installation and execution in most operating systems. To achieve seamless integration, the accomplishment of this work involves the development of two independent modules \textsf{G4LTL} and \textsf{JBernstein}.

\subsection{\textsf{G4LTL}}

\textsf{G4LTL} is the engine that performs (pseudo-Boolean) LTL synthesis. \textsf{G4LTL} implements two (incomplete) algorithms for
synthesizing controllers. The first algorithm translates the corresponding B\"uchi automaton of a given specification to a B\"uchi game and generates the controller via B\"uchi game solving.
The second algorithm parses the B\"uchi automaton of the negated specification, constructs a safety game via bounded unroll, and generates the controller via safety game solving.
Ptolemy~II uses actor-oriented programming techniques, and each actor can be viewed as a function that takes input
tokens and produces output tokens. The engine is designed to allow the environment to start the first move; i.e., no
output is generated if no input is provided\footnote{This read-input-produce-output view is common within the signals and systems community.}.
\textsf{G4LTL} uses two Java-based libraries LTL2Buchi~\cite{Giannakopoulou02fromstates}
and JDD~\cite{jdd}, where LTL2Buchi is used for converting an LTL specification into its corresponding B\"uchi automaton,
and JDD is used for solving safety and B\"uchi games symbolically.

\begin{figure}[t]
	\centering
	\includegraphics[width=\textwidth]{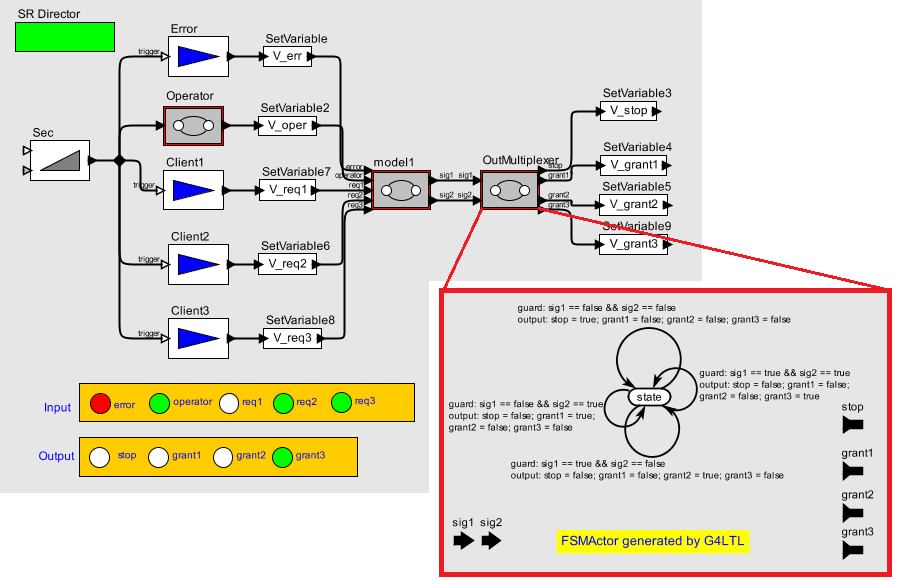}
	\caption{Result of synthesized FSMActor together with the automatically generated multiplexer, given an LTL
    specification in Figure.~\ref{fig:Error.Handling}.}
	\label{fig:PtolemyII.ErrorHandling}
\end{figure}

\textsf{G4LTL} also implements a new feature to automatically rewrite an LTL specification before synthesis (as a preprocessing step),
so that the use of variables can be reduced. This idea was sketched in our earlier technical
report~\cite{cheng:2011:optimization}, and \textsf{G4LTL} is the first tool that realizes this feature for LTL synthesis.
The workflow first synthesizes a controller realizing the rewritten
specification (thus the synthesis speed can be exponentially faster than doing synthesis on the original
specification), followed by automatically producing the signal multiplexer that translates a signal to the original one.
The idea is illustrated in the example in Figure~\ref{fig:Error.Handling}.
For the specification above, \textsf{G4LTL} automatically detects that one can re-encode the output signals, as only~4 output
combinations $\{(1,0,0,0), (0,1,0,0), (0,0,1,0), (0,0,0,1)\}$ are possible. Therefore, it is sufficient to perform
synthesis with two output variables. The system integrated with the synthesized modules is
shown in Figure~\ref{fig:PtolemyII.ErrorHandling}, where \textsf{model1} is an \textsf{FSMActor} realizing the translated specification,
and \textsf{OutMultiplexer} performs signal translation. The integration into Ptolemy~II makes the visualization and simulation
easy. In this model, once when links are established, within the simulation the light signal demonstrates whether a certain variable
is set to \textsf{true}. In Figure~\ref{fig:PtolemyII.ErrorHandling}, although an error appears, as the operator is also present,
the controller can still grant the third client by setting variable \textsf{grant3} to be \textsf{true}.

\subsection{\textsf{JBernstein}}

\textsf{JBernstein} is the engine that handles the validation of (counter-)strategies.
\textsf{JBernstein} implements the validity checking process via the use of Bernstein polynomials.
The process contains three steps: range transformation, basis transformation,
and subspace refinement to recursively derive more precise bounds. As during the design process,
we were unable to use software packages like GNU Multiple Precision Arithmetic Library in C or C++ which can guarantee precision by
representing every rational number with two integer values (e.g., represent
$\frac{1}{3}$ by $\langle 1, 3\rangle$), \textsf{JBernstein} uses \textsf{double} but conservatively estimates
the maximum possible error due to the use of \textsf{double}. In this way, \textsf{JBernstein} only proves or
disproves the validity on a strengthened property (by considering errors) and can also return \textsf{unknown}.
We have used \textsf{JBernstein} to evaluate examples within the NASA PVS benchmark suite.
\textsf{JBernstein} outperforms existing tools on complicated examples,
while the strengthening is still precise enough to provide answer for all problems.

\subsection{Integration}

\paragraph{Constructing pseudo-Boolean game graphs only once.} Our presented algorithm is based on an assumption where we view the synthesis engine and the validity checker as two black boxes. In the implementation, the algorithm can be further improved as we have full control to modify the LTL synthesis engine. Recall in each refinement step caused by the spurious counter-strategy, the algorithm removes the ability of the environment by disallowing some input combinations. It is equivalent to the removal of environment edges in the generated game for the original pseudo-Boolean specification. Therefore, during the integration process, one can modify the internal data structure for games
in \textsf{G4LTL} that marks an edge to be \emph{present} or \emph{absent} (due to the removal).
This process avoids the repeated construction of game arenas from pseudo-Boolean LTL specifications\footnote{The use of CEGAR is an attempt to reduce the number of checks in for the theory solver, as validity checking for nonlinear constraints over reals is very time consuming. However, for the un-optimized method the cost of reducing checks for the theory solver is transferred to the check of realizability for pseudo-Boolean LTL specifications, as every refinement requires to trigger the LTL synthesis engine once.}.

\paragraph{Implementing the \textsf{Extract} module.}
The counter-strategy generation process within the pseudo-Boolean LTL synthesis engine plays an important role for
numerical LTL synthesis. To apply CEGAR techniques, the engine needs to generate a strategy that minimizes the use of
input assignments that are not checked on the theory level (otherwise, the validation can be overly time-consuming).
In the presented counter-strategy, a state may have multiple outgoing edges. If only one edge is chosen for every state in the counter-strategy, it is still a counter-strategy. Therefore, the \textsf{Extract} module first tries to pick for every state in $M_{env}$ an edge whose input element is in $S_{checked}$. If this is possible, then the counter-strategy is genuine. Otherwise, perform a greedy-based approach to cover every state with an edge while maintaining the least use of input assignments that are not checked on the theory level. For the example presented in this paper, \textsf{G4LTL} is able to identify a counter-strategy (from the generated arena) that merely
uses $(\sig{req1}, \sig{req2}) = (\textsf{true}, \textsf{true})$.

\paragraph{Using the tool.}
To use numerical LTL synthesis in Ptolemy~II, currently a user must provide the pseudo-Boolean specification (for \textsf{G4LTL}),
together with polynomial constraint template where each abstract variable is set to \textsf{true} (for \textsf{JBernstein};
\textsf{JBernstein} will automatically parse the template and concretize based on the actual input variable assignment).
The output (whenever a controller exists) is automatically created as an \textsf{FSMActor} component in a design canvas. One can open and see the concrete implementation, link it with other components to complete the design, or even use the code-generation framework
within Ptolemy~II to deploy the synthesized module as executable C program on dedicated platforms in cyber-physical systems.

\begin{figure}[t]
	\centering
	\includegraphics[width=\textwidth]{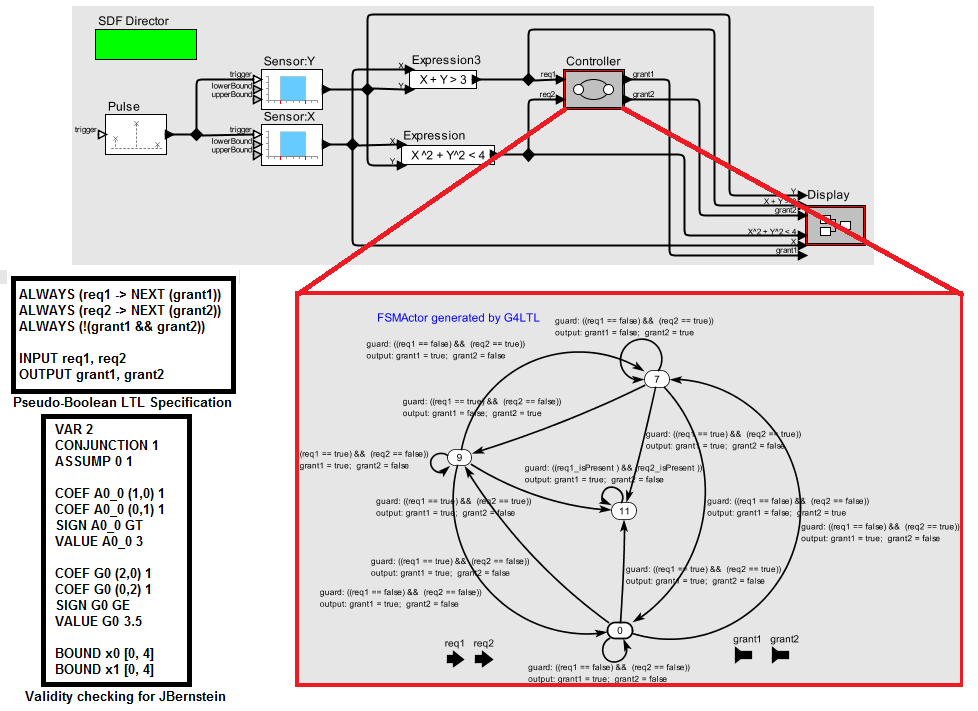}
	\caption{Integrating the synthesized \textsf{FSMActor} into an existing design.}
	\label{fig:PtolemyII}
\end{figure}

Figure~\ref{fig:PtolemyII} shows a Synchronous Dataflow (SDF) model describing the scenario presented in
this paper, together with the synthesized controller (named \textsf{model1}). The \textsf{Purse} actor is
used to generate reactive runs, and the value of sensor \textsf{Sensor:X} and \textsf{Sensor:Y} are
generated at random with a uniform distribution. After data processing, the value of \textsf{$X+Y>3$} is
fed into the input port \textsf{req1} of the synthesized \textsf{FSMActor}, while \textsf{$X^2+Y^2<3.5$}
is fed into the input port \textsf{req2}. Inputs and outputs from the \textsf{FSMActor} are connected to
a console to show the result of simulation. The \textsf{FSMActor} starts with state~\textsf{5}.
Whenever an input (\textsf{true}, \textsf{true}) is provided, the specification holds and the execution
moves to state~\textsf{0}. State~\textsf{7} means that there is a pending grant to be issued for client~2; all outgoing edges (except the one to state~\textsf{0}) set \sig{grant2} to \sig{true}.
Analogously, state~\textsf{9} implies a pending grant to be issued for client~1.

\paragraph{Another example with 3 sensors.} Consider another example where two clients make their requests based on the reading of three sensor values $x_0, x_1, x_2$ ranging within $[0,4]$. Client~1 requests when $x_0+x_1+x_2>3$, while client~2 requests when $(x_0)^2+(x_1)^2+(x_2)^2<4$. The solver detects that there exists a genuine counter-strategy by assigning $(x_0, x_1, x_2)$ to be
$(0.314453125, 1, 1.6875)$. Under this assignment, $x_0+x_1+x_2= 3.001953125>3$ and $(x_0)^2+(x_1)^2+(x_2)^2 = 3.946537017822265625 < 4$.

\section{Concluding Remarks and Discussion}

\paragraph{Availability.}
Currently the tool is within the development version\footnote{Visit \texttt{http://chess.eecs.berkeley.edu/ptexternal/} to download the
experimental version and follow the instructions to execute examples demonstrated in this paper.}\footnote{The \textsf{G4LTL} module in
Ptolemy~II only employs the basic synthesis engine without optimization. The fortiss institute has developed an alternative (academic free) version that contains techniques presented in this paper and the synthesis speed is much faster.}
of Ptolemy~II and is still under active development. Part of the features will be made
public in the release of Ptolemy~II version 9.0. Users can also download separately \textsf{G4LTL}
and \textsf{JBernstein} to experiment the counter-example guided synthesis process manually.

\paragraph{Contribution.}
We conclude this report by summarizing our main contributions.
\begin{itemize}
    \item The proposition of numerical LTL synthesis as a novel way to synthesize controllers for cyber-physical systems.
    \item The proposition of a CEGAR-based algorithm for numerical LTL synthesis, which involves the interplay
        between an LTL synthesizer and a (nonlinear real arithmetic) theory checker.
    \item An implementation within the component-based design tool Ptolemy II that makes both normal and numerical LTL synthesis within the formal method community more approachable to a broader audience.
\end{itemize}

\paragraph{Discussion.}

LTL synthesis has been criticized concerning its practicability for two reasons.
\begin{enumerate}
\item It is very difficult to describe the system specification in full.
\item For full LTL synthesis, existing technologies commonly fails to scale with more than 10 input and 10 output variables.
\end{enumerate}

Nevertheless, within the context of block-based languages such as Ptolemy~II, one can discover that a system is hardly monolithic
but rather composed by several blocks. Each block should be functionally isolated and implements a very specific feature.
It turns out that it is highly uncommon to have the number of input and output ports within a block to be excessively high\footnote{Observe Fig.~\ref{fig:PtolemyII.ErrorHandling}, where for the \textsf{FSMActor} \textsf{model1}, using~5 inputs already occupies the left edge of the block completely.}.
Therefore, complete specification on the block level should be considered as possible.
We believe that understanding the feasibility on block-based languages while probing
further in numerical aspects of control will make synthesis techniques very useful to
the CPS community.


For subsequent work, we plan to extend the technique to use it within the project \textsf{iCyPhy} (industrial cyber-physical
systems)\footnote{\textsf{iCyPhy}: \texttt{http://www.icyphy.org/}} to synthesize over-voltage protection units for electric power systems.

\paragraph{Capturing system dynamics.} For the previous workflow, the validity checking only checks whether a certain pseudo-Boolean input combination is possible on the theory level. For system dynamics captured by an equational system, checking if a counter-strategy is valid also involves a checking whether two (or more) consecutive pseudo-Boolean input valuations are realizable in the theory level.

We also plan to adapt the technique called \emph{relational abstraction}~\cite{DBLP:conf/cav/SankaranarayananT11,DBLP:conf/cav/Tiwari12,DBLP:conf/cav/ZutshiST12}. Given a time frame $\delta$, relational abstraction enables to create a sound approximation concerning relations between inputs for time $t$ and $t+\delta$, for all time $t$. With relational abstraction, our counter-strategy validation process also needs to examine whether all consecutive pseudo-Boolean inputs in the counter-strategy is realizable by being contained inside the relational abstraction. If it is impossible to have $\sig{in}_1$ and $\sig{in}_2$ as two consecutive pseudo-Boolean inputs, then in the refinement process, one does not add a state invariant (as in Figure~\ref{fig:CEGAR}) but an invariant similar in the following form
\begin{equation}
\G \neg( \sig{in}_1 \rightarrow \,\textbf{X}\, \sig{in}_2)
\end{equation}
to act as an assumption on the environment.






\vspace{5mm}
\noindent \textbf{(Acknowledgement)} We thank Christopher Brooks (UC~Berkeley) for his
supports during the tool integration, and Harald Ruess (fortiss GmbH) 
for his initial feedback over this work.

\bibliographystyle{eptcs}

\end{document}